\documentclass[twocolumn, secnumarabic,amssymb, nobibnotes, aps, prb,groupedaddress,superscriptaddress]{revtex4-2}
\usepackage{amsmath,graphicx,latexsym,times,color}
\usepackage{setspace}
\usepackage{hyperref}

\usepackage{array}
\usepackage{textcomp}
\usepackage{titlesec}
\usepackage{physics}
\usepackage{gensymb}
\usepackage{nccmath}
\usepackage{empheq} 
\usepackage[normalem]{ulem} 

\definecolor{blue-violet}{rgb}{0.54, 0.17, 0.89}\newcommand{\V}[1]{\ensuremath{\mathbf{#1}}} 

\let\oldtimes\times  
\renewcommand\times{{\oldtimes}}

\usepackage[cmyk,dvipsnames]{xcolor}
\definecolor{darkorchid}{HTML}{bf3eff}

\begin{document}

  \title{Strongly enhanced lifetime of higher-order bimerons and antibimerons}

\author{Shiwei Zhu}
\affiliation{Universit\'e de Toulouse, CNRS, CEMES, Toulouse, France}
\affiliation{Zhejiang Key Laboratory of Quantum State Control and Optical Field Manipulation, Department of Physics, Zhejiang Sci-Tech University, Hangzhou 310018, China}

\author{Moritz A. Goerzen}
\affiliation{Universit\'e de Toulouse, CNRS, CEMES, Toulouse, France}

\author{Changsheng Song}
\affiliation{Zhejiang Key Laboratory of Quantum State Control and Optical Field Manipulation, Department of Physics, Zhejiang Sci-Tech University, Hangzhou 310018, China}

\author{Stefan Heinze}
\affiliation{Institute of Theoretical Physics and Astrophysics, University of Kiel, Leibnizstrasse 15, 24098 Kiel, Germany}
\affiliation{Kiel Nano, Surface, and Interface Science (KiNSIS), University of Kiel, 24118 Kiel, Germany}
    
\author{Dongzhe Li}
\email[Contact author: ]{dongzhe.li@cemes.fr}
\affiliation{Universit\'e de Toulouse, CNRS, CEMES, Toulouse, France}

	\date{\today}
	
	\begin{abstract}
Magnetic bimerons, similar to skyrmions, are topologically nontrivial spin textures characterized by topological charge $Q$. Most studies so far have focused on low-$Q$ solitons ($|Q| \leq 1$), such as skyrmions, bimerons, and vortices. Here, we present the first calculations of the lifetimes of {ring-like} high-$Q$ bimerons and demonstrate that they are fundamentally more stable than high-$Q$ skyrmions over a wide range of temperature. To obtain realistic results, our chosen system is an experimentally feasible van der Waals interface, Fe$_3$GeTe$_2$/Cr$_2$Ge$_2$Te$_6$. We show that the lifetimes of high-$Q$ (anti)bimerons can exceed the lifetime of those with $|Q|=1$ by 3 orders of magnitude. Remarkably, this trend remains valid even when extrapolated to room temperature (RT), as the lifetimes are dominated by entropy rather than energy barriers. This contrasts with high-$Q$ skyrmions, whose lifetimes fall with $|Q|$ near RT. We attribute this fundamental difference between skyrmions and bimerons to their distinct magnetic texture symmetries, which lead to different entropy-dominated lifetimes. 

\end{abstract}
	
	\maketitle

Magnetic solitons -- localized vortex-like spin structures $\mathbf{m}:\mathbb{R}^2\to\mathbb{S}^2$ with non-trivial topology -- show great promise for spintronic applications \cite{fert2017magnetic,gobel2021beyond}. Their spin textures can be characterized by a topological invariant $Q$, also known as topological charge
\begin{equation}
	Q= \frac{1}{4\pi}\int_{\mathbb{R}^2} \mathbf{m}\cdot\left(\frac{\partial\mathbf{m}}{\partial x_1}\times\frac{\partial\mathbf{m}}{\partial x_2}\right) ~ \mathrm{d}^2\mathbf{r},~~~\mathbf{r} = (x_1, x_2) \in \mathbb{R}^2~,
\end{equation}
{with $\mathbf{m}(\mathbf{r})$ denotes the normalized local magnetization.}
While solitons with the same charge can be homotopically transformed into each other, a transformation with topological phase transition invokes a singular spin configuration, which is usually accompanied by a finite energy barrier \cite{polyakov22metastable}. This leads to the concept of topological protection.  

Over the past decade, solitons with $|Q| \leq 1$, such as skyrmions, skyrmionium, bimerons, and vortices, have been comprehensively studied. Here, {due to their topological similarity, magnetic bimerons, which comprise a meron–antimeron pair, are also considered in-plane skyrmions} \cite{bimeron_PRB2019,Kuchkin2020,Zarzuela2020,Rybakov2025}. These solitons are typically stabilized by the Dzyaloshinskii-Moriya interaction (DMI), and they can be as small as the nanoscale \cite{heinze2011spontaneous,Romming2013,Dongzhe2022_fgt}, move efficiently under current \cite{fert2013,sampaio2013nucleation}, allow topological spin switching by optical pulses (i.e., ultrafast control) \cite{buttner2021observation,dabrowski2022all,khela2023laser}, and can be detected all-electrically \cite{hanneken2015,crum2015perpendicular,li2024proposal}. Recently, the study has progressed to high-$Q$ solitons ($|Q| > 1$) {including isolated high-$Q$ skyrmions \cite{hassan2024dipolar} as well as composite structures} such as skyrmion bundles or skyrmion bags \cite{foster2019two,tang2021magnetic,liu2025room}. 
These high-$Q$ solitons have attracted significant interest due to their
rich physical properties and potential functionalities, such as higher topological complexity \cite{niu2025magnetic} for information encoding and a reduced skyrmion Hall angle \cite{hassan2024dipolar} for efficient information transmission. 

However, the formation mechanisms of high-$Q$ solitons and their stability remain largely unexplored. In particular, the average lifetimes $\tau$ of high-$Q$ solitons, usually described by the Arrhenius law \cite{bessarab2012harmonic}
\begin{equation}\label{tst}
	\begin{split}
		\tau = \Gamma_0^{-1} \exp\left(\frac{\Delta E}{k_\text{B} T}\right)
	\end{split}
\end{equation}
are unknown. The lifetime calculations are challenging due to the complexity of identifying the minimum energy path in a complex energy surface and the need to determine both the energy barrier $\Delta E$ and the entropic contributions incorporated in the pre-exponential factor $\Gamma_0$. Therefore, even for low-$Q$ solitons, lifetimes have been investigated by only a very limited number of groups worldwide \cite{hagemeister2015stability,bessarab2018lifetime,Malottki2019,
	Desplat2020,muckel2021experimental,Dongzhe_PRB2024}. 
In particular, the transition mechanisms and lifetimes of high-$Q$ solitons have not been reported so far.

\begin{figure*}[t]
	\centering
	\includegraphics[width=1.0\linewidth]{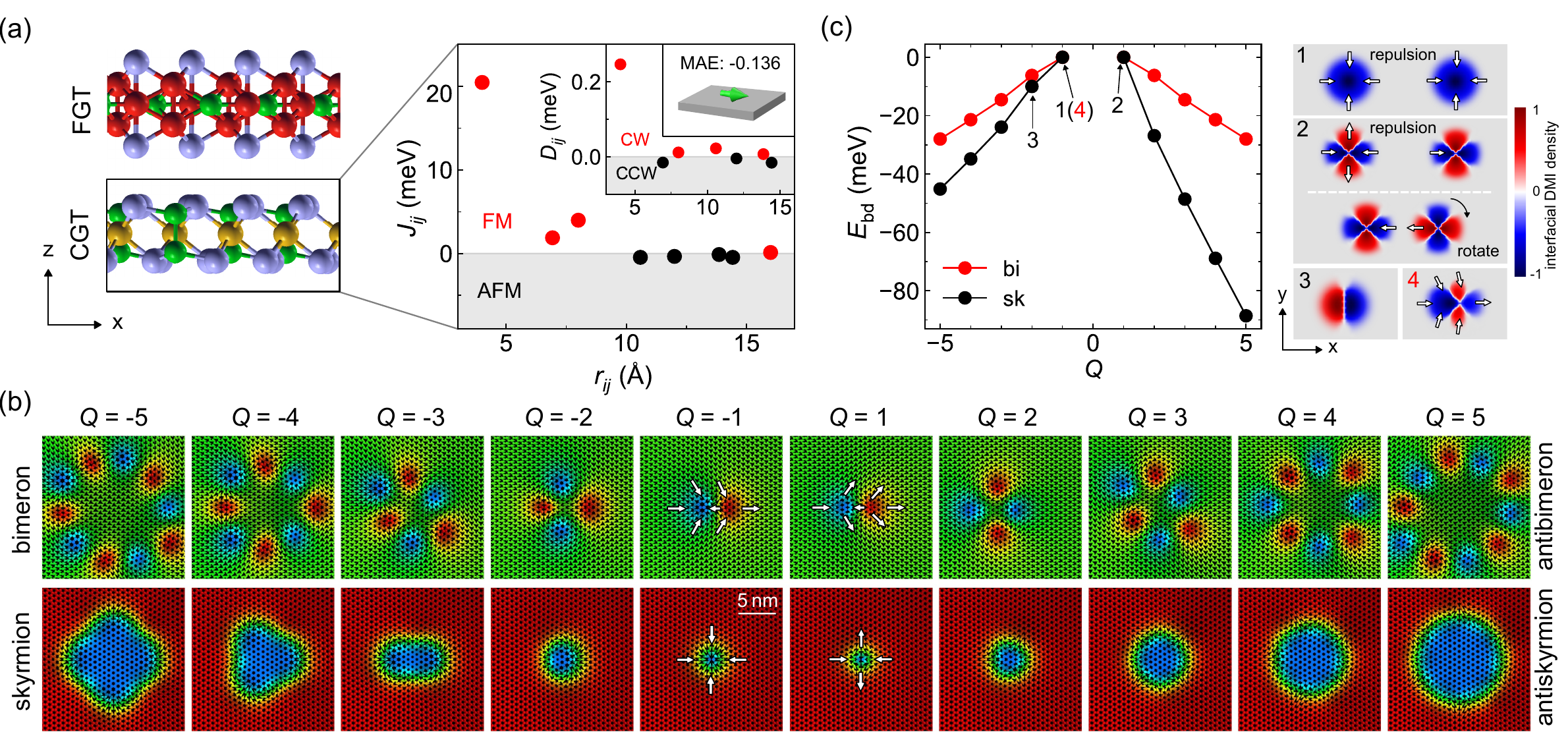}
	\caption{\label{fig_1}Side view of the Fe$_3$GeTe$_2$/Cr$_2$Ge$_2$Te$_6$ (FGT/CGT) van der Waals heterostructure and the calculated exchange interactions and DMI as a function of {distance $r_{ij}$ between sites $i$ and $j$} in arbitrary nearest neighbors of its CGT layer. (b) Top view of relaxed zero-field spin textures with $Q$ ranging from $-5$ to 5 for both bimerons and skyrmions. Scale bar, 5 nm. (c) The binding energy $E_\text{bd}$ of bimerons (bi) and skyrmions (sk) as a function of $Q$. Negative values indicate that high-$Q$ states are energetically more favorable than separated low-$Q$ solitons. The right part displays the interfacial DMI energy densities, defined as
		$\varepsilon_{ij} = (\hat{\mathbf{r}}_{ij} \times \hat{z}) \cdot \left( \mathbf{m}_i \times \mathbf{m}_j \right)$, for low-$Q$ skyrmions (1), antiskyrmions (2), $Q = -2$ skyrmions (3), and low-$Q$ bimerons (4). Red, blue, and white indicate positive, negative, and vanishing Néel DMI energy contributions, respectively, corresponding to Néel-type domain walls with opposite directions and Bloch-type domain walls. For solitons with the same polarity, regions with opposite $\varepsilon$ can bind energy freely.}
\end{figure*}

In this Letter, we present the first calculations of the lifetimes of high-$Q$ solitons, {including isolated high-$Q$ skyrmions and their in-plane ferromagnetic counterparts, namely ring-like high-$Q$ meron bound states referred to here as high-$Q$ bimerons (see Supporting Information (SI) \cite{supplmat} Sec. S2 for structural details).} We demonstrate that high-$Q$ bimerons exhibit significantly longer lifetimes than high-$Q$ skyrmions {under comparable conditions}.
We show the main idea using a van der Waals (vdW) heterostructure, Fe$_3$GeTe$_2$/Cr$_2$Ge$_2$Te$_6$ (FGT/CGT). We predict that high-$Q$ bimerons and antibimerons can coexist at zero field with arbitrary $Q$. Combining first-principles calculations with transition state theory, we further demonstrate that their lifetimes are enhanced by several orders of magnitude compared to those of low-$Q$ counterparts over a wide temperature range, even at room temperature (RT). In contrast, the lifetimes of high-$Q$ skyrmions are shorter than those of their low-$Q$ counterparts near RT. We argue that this contrast arises from entropy-driven stability, which is governed by the symmetries of their magnetic textures.

We consider an all-magnetic vdW heterostructure composed of FGT and CGT (see Fig.~\ref{fig_1}(a)), which was recently synthesized experimentally \cite{wu2022van}. In this work, we focus on solitons in the CGT layer, which exhibit a delicate balance between exchange frustration, {characterized by the competition between ferromagnetic (FM) and antiferromagnetic (AFM) couplings that can support topological solitons \cite{von2017enhanced, rybakov2022magnetic}, }and DMI, combined with in-plane magnetocrystalline anisotropy energy (MAE), as discussed below. 
In CGT, the Cr atoms form a two-dimensional honeycomb lattice 
comprising two Cr atoms per unit cell
(see Fig.~S1 in SI \cite{supplmat}). To investigate magnetic interactions of the CGT layer, we map our DFT total energies to the following spin Hamiltonian for Cr atoms:
\begin{equation}\label{hamiltonian}
	H = -\sum_{ij} J_{ij} \, (\mathbf{m}_i \cdot \mathbf{m}_j) 
	- \sum_{ij} \mathbf{D}_{ij} \cdot (\mathbf{m}_i \times \mathbf{m}_j)
	- K \sum_i (m_i^z)^2,
\end{equation}
where $\V{m}_i$ and $\V{m}_j$ are unit magnetic moment vectors at lattice sites $i$ and $j$. The three terms denote Heisenberg exchange ($J_{ij}$), DMI ($\V{D}_{ij}$), and MAE ($K$). Dipole-dipole interactions are not included in our model. Since the CGT layer is a monolayer system, {the long-range dipolar contribution is significantly weaker than the exchange and anisotropy interactions (see SI Sec. S4 for details).}

Our density functional theory (DFT) calculations predict that CGT exhibits exchange frustration, in which {short-range FM and long-range AFM interactions} compete with each other (Fig.~\ref{fig_1}(a), for computational details, see SI Sec.~S1 \cite{supplmat}). Additionally, the DMI arises from inversion symmetry breaking and spin-orbit coupling induced by the adjacent FGT layer, which makes it promising to study the interplay between these two interactions in stabilizing higher-order topological spin textures. Finally, the in-plane MAE of CGT favors bimerons as metastable configurations.

\begin{figure*}[t]
	\centering
	\includegraphics[width=1.0\linewidth]{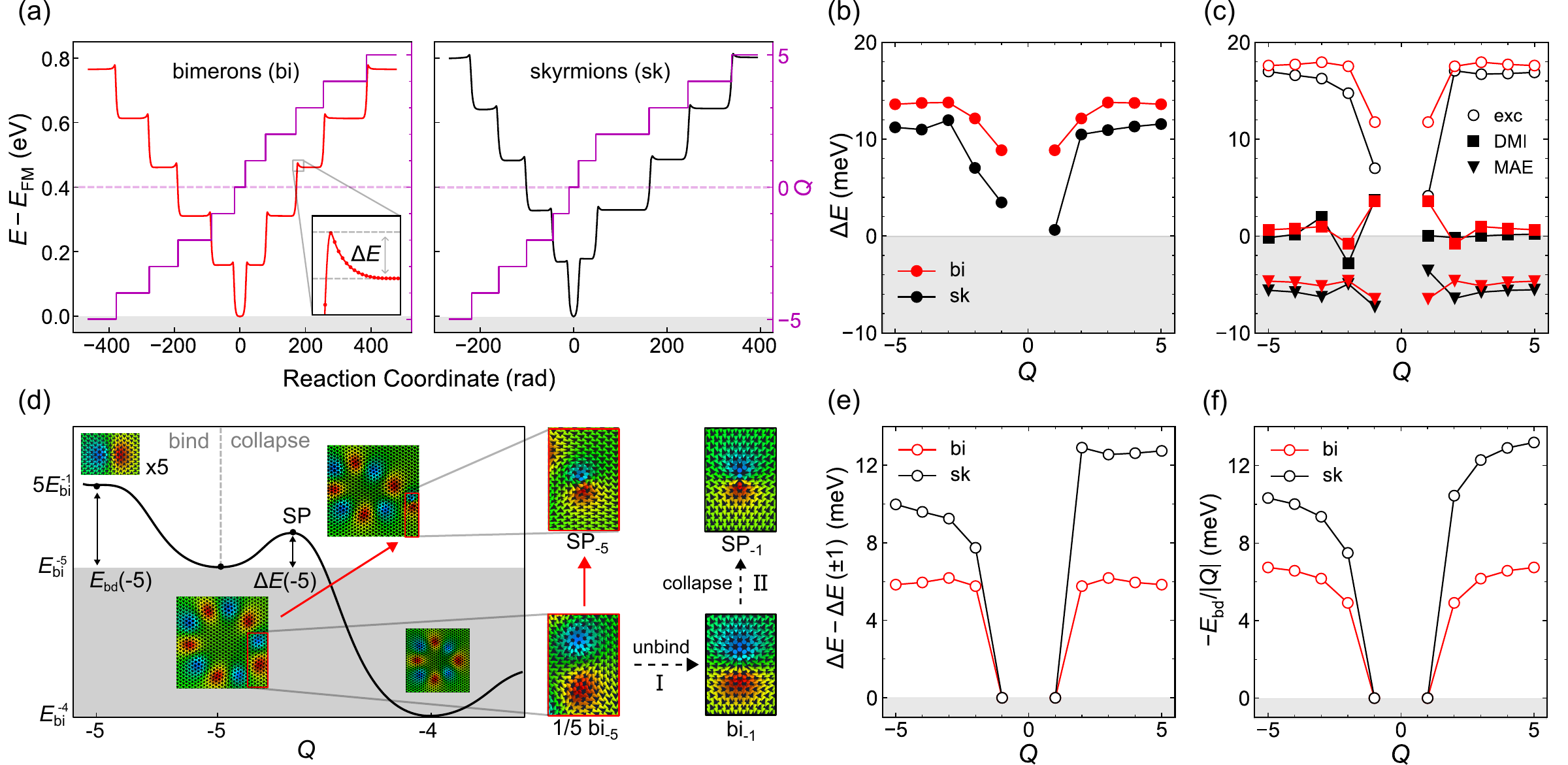}
	\caption{\label{fig_2}(a) Minimum energy path (MEP) of bimerons (left) and skyrmions (right) with $Q$ ranging from $-5$ to $5$. {The reaction coordinate is defined as the cumulative geodesic distance between successive spin configurations along the MEP, measured on the unit sphere for each spin.} The inset shows the definition of energy barrier ($\Delta E$), corresponding to stepwise transitions between neighboring topological charges. (b) Total energy barriers for bimerons (red) and skyrmions (black) as a function of $Q$. (c) Same as (a), but showing the energy decomposition: exchange (open circles), DMI (filled squares), and MAE (filled triangles). (d) Schematic illustration of the energy curve during the binding of a $Q = -5$ bimeron and its collapse to a $Q = -4$ bimeron. The inset shows the spin configurations of a $Q = -1$ bimeron, the $Q = -5$ bimeron and its SP, and the $Q = -4$ bimeron. The energy barrier mainly comes from the collapse of one meron pair, as indicated by the red arrow and rectangles. The right part of (d) shows its approximation with two subprocesses. The exchange energy contributions for $\Delta E - \Delta E_{\text{II}}$ (e) and $E_{\text{I}}$ (f) are compared to verify this approximation, corresponding to the energy barrier difference and the binding energy, respectively. The consistency between these two panels indicates that the increase in the exchange energy barrier for high-$Q$ states originates from the unbinding process (I).
	}
\end{figure*}

Our spin dynamics simulations with the spin Hamiltonian parameterized by first-principles predict the emergence of nanoscale bimerons ($Q<0$) and antibimerons ($Q>0$)
for arbitrary $Q$ (Fig.~\ref{fig_1}(b), top). We used $80 \times 80$ lattices (2-atom per unit cell) in this work to obtain well-converged results. As reported in Ref. \cite{Moritz_bimeron2025}, bimerons and antibimerons have long-range spatial profiles. Interestingly, we find that high-$Q$ bimerons are much more localized, which can be related to them always forming ring-like structures \cite{Zhang2020} to minimize their energy (see SI Sec.~S3 \cite{supplmat}). For fair comparison with skyrmions, we switch the MAE to out-of-plane (Fig.~\ref{fig_1}(b), bottom), i.e.~changing only the sign of $K$. {In addition, a slight adjustment to the interactions is required to stabilize $Q = 1$ antiskyrmions (see SI Sec.~S4 \cite{supplmat}), since they rely on frustrated exchange. However, this does not affect any of our main conclusions (see Sec.~S11 for a robustness analysis).} Field-induced bimeron–skyrmion conversion occurs at low $Q$ \cite{Moritz_bimeron2025} but fails at high $Q$ due to large Zeeman penalties (SI Sec.~S5 \cite{supplmat}). These bimerons appear symmetric to their antibimeron counterparts, whereas skyrmions are asymmetric with antiskyrmions. As we will demonstrate in the following, one important aspect of our study is to understand the fundamental difference between high-$Q$ bimerons and skyrmions.

To understand high-$Q$ soliton formation, we compute the binding energy $E_\text{bd} = E(Q)-|Q|E(\pm1)$, describing the energy gain when $|Q|$ primitive solitons with charge $\pm1$ combine. $E_\text{bd}$ is symmetric for bimerons vs.~antibimerons but asymmetric for skyrmions vs.~antiskyrmions (Fig.~\ref{fig_1}(c)), and its negative sign indicates that high-$Q$ states are favored. Often, the binding is barrierless (SI Sec.~S6 \cite{supplmat}). The interfacial DMI density $\varepsilon_{ij} = (\hat{\mathbf{r}}_{ij} \times \hat{z}) \cdot (\mathbf{m}_i \times \mathbf{m}_j)$ (arrows show in-plane $\mathbf{m}_{i,\parallel}$) reveals that skyrmions have uniform $\varepsilon$, whereas antiskyrmions show $\varepsilon<0$ along $x$ and $\varepsilon>0$ along $y$. Solitons with regions of opposite $\varepsilon$ can rotate to facilitate barrierless binding \cite{sk_bind_ask, ask_bind_ask} (Animation 1, SI \cite{supplmat}), including high-$Q$ skyrmions and bimerons. High-$Q$ states thus form spontaneously upon close approach, except for repulsive $Q=-1$ skyrmions.

To quantify the thermal stability of solitons, we compute lifetimes using harmonic transition state theory (HTST) \cite{bessarab2012harmonic}, {which simplifies the complex energy landscape by applying a harmonic approximation to the potential energy surface near the local energy minima and the saddle points (SPs). By evaluating the frequency of thermal attempts and the probability of overcoming the energy barrier, HTST allows for the direct calculation of the transition rate and the corresponding lifetime} (see SI Sec.~S1 for theory and computational details). Since the binding process is generally barrierless, lifetimes are solely determined by collapses. 
In our case, the collapse of a high-$Q$ soliton into the FM ground state typically proceeds through a series of step-by-step transitions, as shown in Fig.~\ref{fig_2}(a). {Consequently, the reported lifetime corresponds to the transition from a state with absolute topological charge $|Q|$ to the adjacent state with $|Q|-1$. The width of the reaction coordinate (cumulative geodesic distance) reflects the underlying dynamics, highlighting the similar collapse processes of bimerons and antibimerons versus the distinct pathways of skyrmions and antiskyrmions.}

Within the harmonic approximation, the Arrhenius law (Eq.~(\ref{tst})) gives the lifetime in terms of the $\Delta E$ and $\Gamma_0$.
The energy barriers $\Delta E$ between neighboring topological charges (inset of Fig.~\ref{fig_2}(a)) are shown in Fig.~\ref{fig_2}(b) for all transitions $Q \to Q'$. Here, $\Delta E(Q)$ increases with $|Q|$ and saturates near $|Q| = 3$ for both bimerons and skyrmions. As shown in Fig.~\ref{fig_2}(c), this behavior stems {mainly} from the exchange interaction, which dominates the stability of all solitons. The $Q$-dependent exchange energy contribution can be understood via two subprocesses (Fig.~\ref{fig_2}(d)):
(I) unbinding a single (anti-)bimeron from the ring structure and (II) its subsequent collapse {of an isolated, primitive soliton with charge $\pm1$}. The unbinding costs an equal share of the total binding energy, $\Delta E_{\text{I}} = -E_{\text{bd}}(Q)/|Q|$, while the collapse energy {$\Delta E_{\text{II}}=\Delta E(\pm1)$} is shown in Fig.~\ref{fig_2}(b) for $Q=\pm1$.
Our {phenomenological} energy barrier model $\Delta E(Q) \approx \Delta E_{\text{I}}(Q) + \Delta E_{\text{II}}$ 
then simply reads
\begin{equation}\label{bd-eb}
	\begin{split}
		\Delta E(Q)\approx \Delta E(\pm1) - \frac{E_{\text{bd}}(Q)}{|Q|}~,
	\end{split}
\end{equation}
indicating that the increased energy barriers for high-$Q$ solitons originate from binding energy {and saturates for sufficiently large $|Q|$}. 

This model allows for numerical verification by comparing the energy barrier differences $\Delta E(Q)-\Delta E(\pm1)$ (Fig.~\ref{fig_2}(e)) with the binding energy (Fig.~\ref{fig_2}(f)), considering only the exchange interaction. The remarkable agreement confirms the effectiveness of Eq.~(\ref{bd-eb}) for high-$Q$ barriers.  {Consequently, the gradual saturation of the binding energy per $Q$ naturally results in a plateauing trend for the energy barriers as $|Q|$ increases. However, deviations occur at low $|Q|$, as the approximation assumes that the collapse of a meron pair does not affect the remaining texture. This assumption is less accurate at low $|Q|$, where the annihilating meron pair constitutes a substantial fraction of the entire texture, thus having a stronger impact on the remaining spins.} 

Secondly, the DMI specifically stabilizes $Q=-1$ skyrmions, while its influence is significantly diminished for skyrmions with $Q \neq -1$~\cite{nagaosa2013} (see SI Sec.~S8 \cite{supplmat}). Consequently, the DMI effectively raises the barrier for $Q=-1$ but lowers it for $Q=-2$ skyrmions, as $Q=-1$ states have lower DMI energy and can form via the collapse of $Q=-2$ skyrmions. Similar but weaker effects occur for bimerons and antibimerons, which also feature DMI-favored regions. {As $|Q|$ increases beyond 2, this chiral modulation becomes relatively less significant, causing the barriers to be governed primarily by the exchange interaction, which leads to the observed saturation.} Besides, the MAE plays a minor role, reducing $\Delta E$ by a nearly constant amount for $|Q|>2$ solitons.

\begin{figure}[t]
	\centering
	\includegraphics[width=1.0\columnwidth]{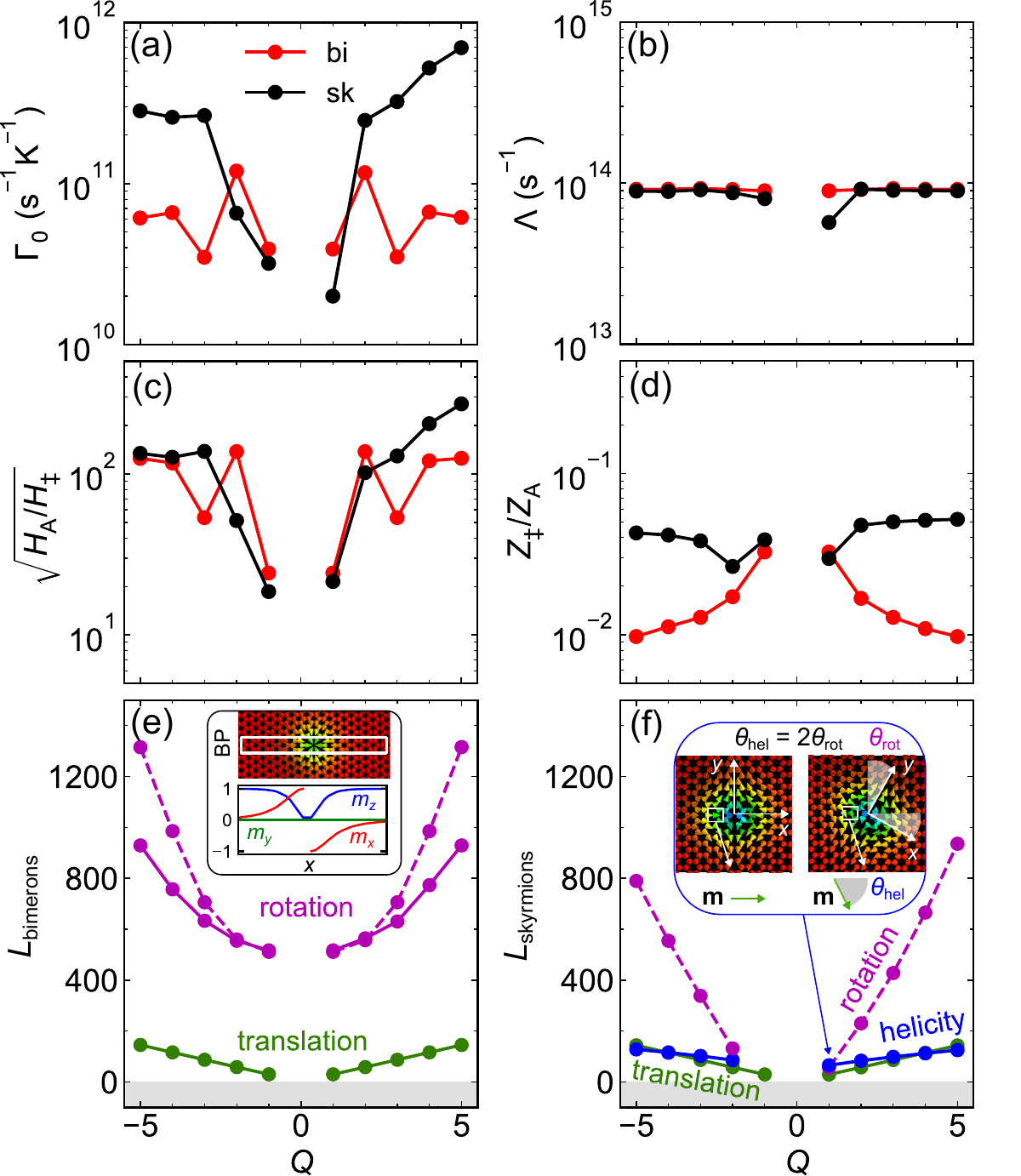}
	\caption{\label{fig_3} 
		Dependence of (a) the pre-exponential factor $\Gamma_{0}$ (with $T = 1 \,\text{K}$,  cf.~Eq.~(\ref{pf}), (b) the dynamical prefactor $\Lambda$, (c) the contribution of nonzero modes to $\Gamma_{0}$ within the harmonic approximation, $\sqrt{H_{\mathrm{A}}/H_{\mathrm{\ddagger}}}$, and (d) the contribution of zero modes to $\Gamma_{0}$, ${Z_{\mathrm{\ddagger}}/Z_{\mathrm{A}}}$, on the topological charge $Q$ for bimerons and skyrmions. (e-f) Characteristic lengths of translational (green), rotational (purple), and helicity (blue) modes for bimerons and skyrmions, respectively. Solid and dotted lines correspond to the initial and SP states, respectively. {The inset in (e) shows a Bloch-point-like defect (BP) at the skyrmion ($Q = -1$) saddle point; the spin components along the cross-section (indicated by the white box) are plotted below, showing a clear discontinuity in $m_x$ at the defect center.} The inset in (f) displays the helicity mode of antiskyrmions ($Q = 1$), characterized by $\Delta \gamma = \theta_{\text{hel}}$. This mode is degenerate with the rotational mode only in the initial state, where the symmetric magnetic textures impose the relation $\theta_{\text{hel}} = 2\theta_{\text{rot}}$, but it becomes a nonzero mode at the SP.
	} 
\end{figure}

{In addition to the energy barrier, the soliton lifetime is also determined by the activation entropy $\Delta S$ and dynamical aspects of the collapse mechanism, which, within HTST, the Arrhenius pre-exponential factor $\Gamma_0$ is expressed as:
\begin{equation}
	\label{pf}
	\begin{aligned}
		\Gamma_0 &= \frac{\Lambda}{2\pi}
		\exp\left( \frac{\Delta S}{k_{\mathrm{B}}} \right), \\
		\Delta S &= k_{\mathrm{B}} \ln \Bigg[
		\left(2\pi k_{\mathrm{B}}T\right)^{\frac{k_{\mathrm{A}}-k_{\ddagger}}{2}}
		\frac{\prod_{i=2}^{1+k_{\ddagger}} L_i^{\ddagger}}
		{\prod_{i=1}^{k_{\mathrm{A}}}L_i^{\mathrm{A}}}
		\times
		\sqrt{
			\frac{\prod_{n=1+k_{\mathrm{A}}}^{2N}\lambda_n^{\mathrm{A}}}
			{\prod_{n=2+k_{\ddagger}}^{2N}\lambda_n^{\ddagger}}
		}
		\Bigg] .
	\end{aligned}
\end{equation}
{here, $\Lambda$ is the dynamical prefactor derived from the Landau–Lifshitz equation, characterizing the system's evolution speed at the SP. 
	The terms $\lambda_n > 0$ denote the non-zero Hessian eigenvalues, corresponding to harmonic modes and representing the curvatures of the energy landscape. Specifically, a relatively flat energy surface at the initial state (superscript ``A'') combined with a sharply curved surface at the saddle point (superscript ``$\ddagger$'') yields a smaller entropy contribution to $\Gamma_0$, thereby enhancing the stability. 
	Furthermore, $k$ denotes the number of zero modes, which correspond to degrees of freedom that preserve the energy (such as global translations or rotations), while $L_i$ are the characteristic lengths associated with these trajectories in the configuration space.}

We plot in Fig.~\ref{fig_3}(a) the dependence of $\Gamma_0$ on $Q$. 
The bimeron with $Q=\pm3$ exhibits a smaller $\Gamma_0$ than low-$Q$ bimerons, indicating enhanced stability, while high-$Q$ skyrmions remain entropically unfavored, as reflected by higher $\Gamma_0$.
To elucidate the origin of this behavior, we decompose the main contributing terms in Eq.~(\ref{pf}) into the dynamical prefactor, the harmonic mode within the harmonic approximation, and the zero-mode. Their respective dependences on $Q$ are displayed in Fig.~\ref{fig_3}(b-d). The values of $\Lambda$ remain nearly constant across different solitons, suggesting that its impact on the $Q$-dependence of lifetimes is minor.

The harmonic contribution $\sqrt{H_{\mathrm{A}}/H_{\mathrm{\ddagger}}}$, with $H_{\mathrm{A}}=\prod_{n=1+k_{\text{A}}}^{2N}\lambda_n^{\text{A}}$ and $H_{\mathrm{\ddagger}}=\prod_{n=2+k_{\ddagger}}^{2N}\lambda_n^{\ddagger}$, generally leads to a larger $\Gamma_0$ for high-$Q$ states (Fig.~\ref{fig_3}(c)). 
The only exception to this rule is bimeron states with $Q=\pm3$. For these states, we find an exceptionally low eigenvalue for the 3-fold distortion mode ({see SI Sec.~S10 \cite{supplmat}}), which is responsible for the comparably low harmonic contribution. Arising from coinciding lattice- and soliton geometry, this means that bimerons with $Q=\pm3$ are entropically stabilized by the $C_3$-symmetry of the honeycomb lattice. However, overall, the harmonic term disfavors high-$Q$ states.

In contrast, the zero-mode contribution ${Z_{\mathrm{\ddagger}}/Z_{\mathrm{A}}}$, with $Z_{\mathrm{\ddagger}}=\prod_{i=2}^{1+k_{\ddagger}}L_i^{\ddagger}$ and $Z_{\mathrm{A}}=\prod_{i=1}^{k_{\text{A}}}L_i^{\text{A}}$, effectively reduces $\Gamma_0$ for high-$Q$ bimerons, while it shows only a minor $Q$-dependence for skyrmions (Fig.~\ref{fig_3}(d)).
To clarify this difference, we analyze the zero mode contribution in the following.

Typically, a 2D soliton possesses two translational zero modes (tra) and, if rotational symmetry is broken, additional rotational zero modes (rot)
\cite{goerzen2023lifetime, Moritz_bimeron2025}.
In this regard, the $Q = -1$ skyrmion is special, because the rotational mode vanishes due to its radial symmetry. For skyrmions and antiskyrmions with $Q\neq-1$, the length $L_{\text{rot}}$ of a single rotation around the soliton center, at radial distance $\rho=0$, is directly linked to a rotation of its helicity with $L_{\text{hel}}$ by
\begin{equation}\label{eq:relation_rot_hel_initial}
	L_{\text{rot}}(0) = |Q+1|L_{\text{hel}} ~,
\end{equation}
as illustrated in the inset of Fig.~\ref{fig_3}(f)  (see SI Sec.~S10 \cite{supplmat} for details).

However, this relation does not hold at the SP. In this work, all SPs exhibit a Bloch-point-like defect (BP), which serves as the {2D lattice counterpart of a Bloch-point in continuum models \cite{pylypovskyi2012bloch, PhysRevLett.98.117201, milde2013unwinding, doring1968point}. As illustrated in the inset of Fig.~\ref{fig_3}(e), this defect is characterized by a central singularity where the spin components (e.g., $m_x$) exhibit a sharp discontinuity.} 
At this localized defect, the topological charge changes abruptly by an integer during the topological transition.
The motion of the BP incurs a substantial energy cost on the honeycomb lattice, rendering the translational mode nonzero, and its contribution $L_{\text{tra}}^{\text{A}}$ appear only in the denominator of Eq.~(\ref{pf}), thereby reducing $\Gamma_0$ and enhancing entropic stabilization at large $Q$ (Fig.~\ref{fig_3}(e–f)).
Moreover, high-$Q$ solitons tend to collapse via a chimera-like pathway \cite{muckel2021experimental, desplat2019paths} (see Animations 4 and 5, and Sec. S9 of the SI \cite{supplmat}), which breaks rotational symmetry and thereby makes the helicity mode nonzero. In this situation, we find that the length $L_{\text{rot}}(\text{BP})$ of a rotation around the BP, at distance $\rho_{\text{BP}}$ from the soliton center, always holds (SI Sec. S1 \cite{supplmat})
\begin{equation}\label{ratio}
	L_{\text{rot}}^{\ddagger}(0) < L_{\text{rot}}^{\ddagger}(\text{BP}) \lesssim L_{\text{rot}}^{\ddagger}(0) + 2\pi\rho_{\text{BP}} L_{\text{tra}}^{\ddagger} ~.
\end{equation}

In combination with Eq.~(\ref{eq:relation_rot_hel_initial}), this relation states the ratio ${L_{\mathrm{rot}}^{\ddagger}}(\text{BP})/L_{\mathrm{hel}}^{\mathrm{A}} \gtrsim |Q+1|$ for skyrmions, assuming that A and SP do not significantly differ in size. Therefore, the rotation modes increase $\Gamma_0$ for skyrmions, counteracting the stabilization by the contribution of translation modes. For bimerons however, the fundamental asymmetry from Eq.~(\ref{eq:relation_rot_hel_initial}) does not apply (cf. Fig.~\ref{fig_3}(e)), leaving a ratio ${L_{\mathrm{rot}}^{\ddagger}}(\text{BP})/L_{\mathrm{rot}}^{\mathrm{A}} \approx 1$. Overall, higher-$Q$ solitons tend to exhibit larger entropic contributions from nonzero and rotational modes due to their increased size. Deviations from this trend mainly arise from translational modes present only in the initial state, as well as symmetry-related effects discussed above. Therefore, high-$Q$ bimerons are more entropically favored than high-$Q$ skyrmions, due to their distinct magnetic texture symmetries. 

\begin{figure}[t]
	\centering
	\includegraphics[width=1.0\columnwidth]{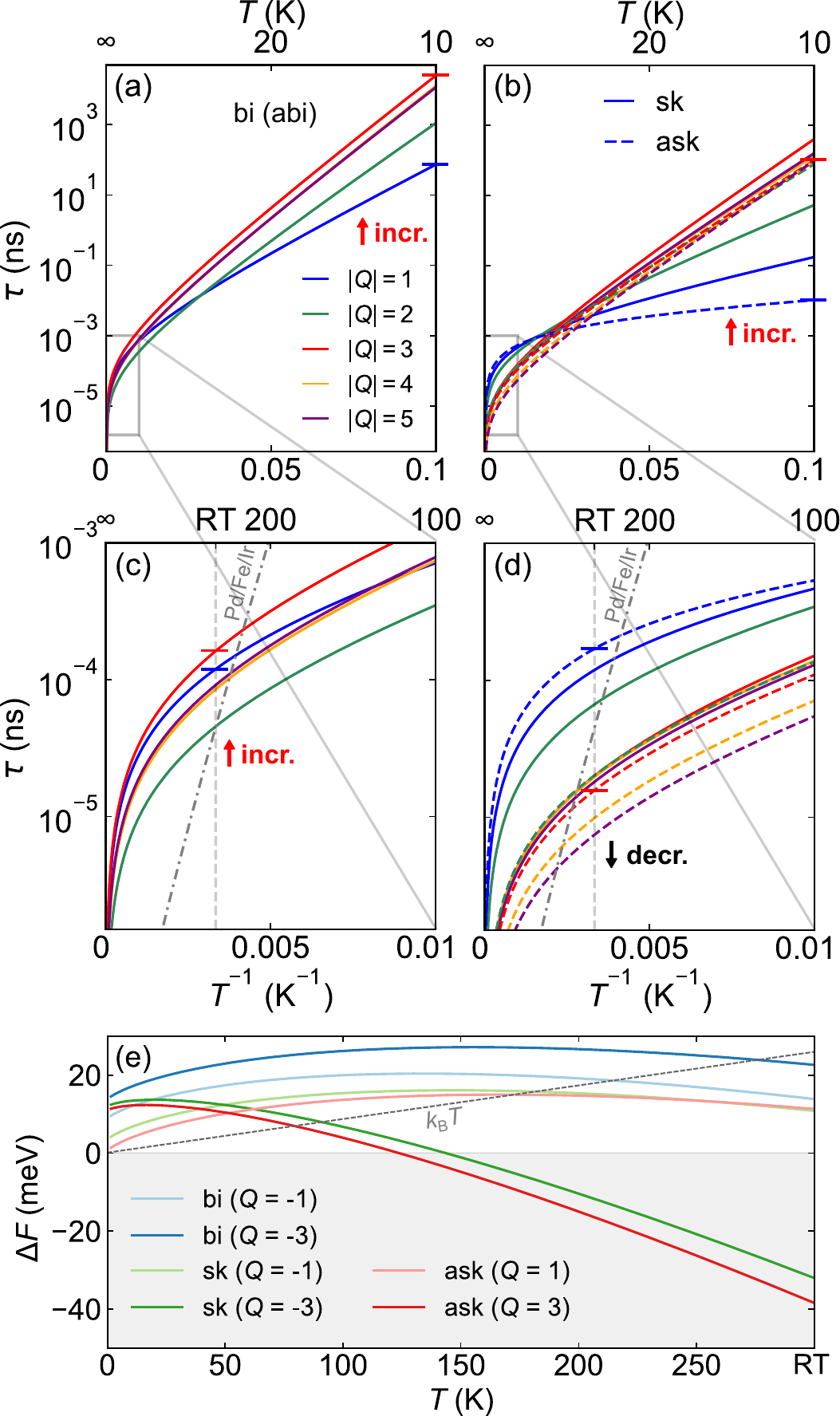}
	\caption{\label{fig_4} (a) Lifetimes of bimerons (bi) and antibimerons (abi) at $B = 0$ T shown over inverse temperature for different $Q$. (b) Same as (a) but for skyrmions (sk) and antiskyrmions (ask). (c-d) corresponding high-temperature ($T = 300$ K) behavior, zoomed in from (a) and (b), respectively. The markers highlight the lifetimes of $|Q|=1$ (blue) and $|Q|=3$ (red) bimerons and antiskyrmions at 10 K and room temperature (RT). For comparison, the skyrmion lifetime in Pd/Fe/Ir(111) at $B = 3.9$ T (dashed-dotted lines) is given \cite{Malottki2019}.
		{(e) Calculated free energy barrier $\Delta F(T) = \Delta E - T\Delta S$ as a function of temperature for selected bimerons, skyrmions, and antiskyrmions. The slopes of the curves illustrate the impact of activation entropy $\Delta S$ on the thermal stability of different topological solitons.}
	}
\end{figure}

Having established $\Delta E$ and $\Gamma_0$, we plot in Fig.~\ref{fig_4}(a-b) the logarithmic Arrhenius law, $\ln \tau = \beta \Delta E - \ln \Gamma_0$, with $\beta^{-1} = k_{\text{B}} T$, for different values of $Q$.
At low temperature ($T \approx 10~\mathrm{K}$), $\tau$ increases with $|Q|$ for both bimerons and skyrmions, rising by around 3 orders of magnitude for bimerons ($|Q|=1$ to 3) and up to 4 for antiskyrmions. This enhancement of the lifetime at low temperatures is governed by $\Delta E$. However, entropy becomes the dominant factor when extrapolated to room temperature (RT). 
{To elucidate this transition, we evaluate the temperature-dependent free energy barrier $\Delta F(T) = \Delta E - T\Delta S$ in Fig.~\ref{fig_4}(e). While both bimerons and skyrmions exhibit an increased $\Delta E$ as $|Q|$ grows, their thermal stability is distinguished by the activation entropy $\Delta S$, which is reflected in the negative slope of the $\Delta F$ curves. For skyrmions and antiskyrmions, the slope increases significantly from $|Q|=1$ to $3$, causing A negative $\Delta F$ at RT. In contrast, high-$Q$ bimerons maintain a relatively flat slope, comparable to their $|Q|=1$ counterparts.}

{Consequently, as shown in Fig.~\ref{fig_4}(c), $\tau$ for $|Q|=3$ bimerons remains higher than that of low-$Q$ counterparts near RT, whereas skyrmion lifetimes decrease significantly with $|Q|$ (Fig.~\ref{fig_4}(d)).}  For comparison, we also plot the lifetime of Pd/Fe/Ir(111) \cite{Malottki2019}, an ultrathin-film system known for hosting nanoscale skyrmions with long lifetimes \cite{Romming2013,muckel2021experimental}. Surprisingly, owing to the relatively flat $\tau$–$T$ curves, the $|Q|=1$ solitons in our material exhibit even longer lifetimes than those in Pd/Fe/Ir(111) at RT. In particular, for high-$Q$ bimerons, {the lifetime can be further enhanced due to their increased $\Delta E$}, highlighting their potential for device applications. We note that as $\Delta F$ approaches the scale of $k_{\text{B}}T$ (dashed line in Fig.~\ref{fig_4}(e)), the harmonic approximation reaches its theoretical limit due to increased anharmonic fluctuations and recrossing events \cite{hanggi1990reaction}. Therefore, the HTST may overestimate the absolute lifetimes; however, this does not affect our core conclusion regarding the opposite trends between skyrmions and bimerons, which are fundamentally governed by their distinct magnetic texture symmetries.

In summary, we propose that {ring-like} high-$Q$ bimerons exhibit a strongly enhanced lifetime compared to high-$Q$ skyrmions {under comparable conditions}, which could make them attractive candidates for applications requiring thermal stability. To obtain realistic results, we chose an experimentally feasible FGT/CGT interface as our model system, in which we demonstrate the coexistence of high-$Q$ bimerons and antibimerons with arbitrary $Q$. We further show that the lifetimes of high-$Q$ bimerons differ fundamentally from those of high-$Q$ skyrmions due to differences in their distinct rotational symmetries, leading to enhanced entropic stability. In combination with the tendency of high-$Q$ bimerons to form superstructures exhibiting a high degree of nonlinear soliton-soliton interactions, we suggest that they may be of interest for neuromorphic and stochastic spintronic devices \cite{song2020} or quantum computing \cite{skyquibit_2021}. 
Our predictions can be experimentally tested using, for instance, spin-polarized scanning tunneling microscopy \cite{muckel2021experimental} or magnetic force microscopy \cite{koraltan2025signatures}.


\textbf{Acknowledgement:}  This study has been supported through the ANR Grant No. ANR-22-CE24-0019. This work is supported by France 2030 government investment plan managed by the French National Research Agency under grant reference PEPR SPIN – [SPINTHEORY] ANR-22-EXSP-0009. This study has been (partially) supported through the grant NanoX no.~ANR-17-EURE-0009 in the framework of the "Programme des Investissements d’Avenir". This work was performed using HPC resources from CALMIP (Grant No. 2023/2025-[P21023]). We thank H. Schrautzer for helpful discussions. 
	
	
\bibliography{References}

\end{document}